\documentclass{article}

\usepackage{microtype}
\usepackage{graphicx}
\usepackage{subfigure}
\usepackage{booktabs} % for professional tables

\usepackage{hyperref}

\usepackage[accepted]{icml2023}

\usepackage{amsmath}
\usepackage{amssymb}
\usepackage{mathtools}
\usepackage{amsthm}

\usepackage[capitalize,noabbrev]{cleveref}
\usepackage{astjnlabbrev}

\theoremstyle{plain}

\theoremstyle{definition}

\theoremstyle{remark}

\usepackage[textsize=tiny]{todonotes}

\icmltitlerunning{Spatio-Spectroscopic Anomaly Detection in MaNGA IFS Data}

\defcitealias{Comerford21}{C20}

\begin{document}

\twocolumn[
\icmltitle{Spatio-Spectroscopic Representation Learning using  \\ Unsupervised Convolutional Long-Short Term Memory Networks}



\icmlsetsymbol{equal}{*}

\begin{icmlauthorlist}
\icmlauthor{Kameswara Bharadwaj Mantha}{min}
\icmlauthor{Lucy Fortson}{min}
\icmlauthor{Ramanakumar Sankar}{min}
\icmlauthor{Claudia Scarlata}{min}
\icmlauthor{Chris Lintott}{ox}
\icmlauthor{Sandor Kruk}{esa}
\icmlauthor{Mike Walmsley}{mcr}
\icmlauthor{Hugh Dickinson}{op}
\icmlauthor{Karen Masters}{hav}
\icmlauthor{Brooke Simmons}{lan}
\icmlauthor{Rebecca Smethurst}{ox}
\end{icmlauthorlist}

\icmlaffiliation{min}{Department of Physics and Astronomy, University of Minnesota Twin Cities, Minneapolis, USA}
\icmlaffiliation{ox}{Company Name, Location, Country}
\icmlaffiliation{mcr}{Jodrell Bank Centre for Astrophyics, Department of Physics and Astronomy, University of Manchester, Manchester, United Kingdom}
\icmlaffiliation{lan}{Physics Department, Lancaster University, Lancaster, LA1 4YB, UK}
\icmlaffiliation{esa}{European Space Agency (ESA), European Space Astronomy Centre (ESAC), Camino Bajo del Castillo s/n 28692 Villanueva de la Cañada, Madrid, Spain}
\icmlaffiliation{hav}{Departments of Physics and Astronomy, Haverford College, 370 Lancaster Avenue, Haverford, Pennsylvania 19041, USA}
\icmlaffiliation{op}{School of Physical Sciences, The Open University, Milton Keynes, MK7 6AA, UK}

\icmlcorrespondingauthor{Kameswara Bharadwaj Mantha}{manth145@umn.edu}

\icmlkeywords{Machine Learning, ICML}

\vskip 0.3in
]



\printAffiliationsAndNotice{} % otherwise use the standard text.

\begin{abstract}
Integral Field Spectroscopy (IFS) surveys offer a unique new landscape in which to learn in both spatial and spectroscopic dimensions and could help uncover previously unknown insights into galaxy evolution. In this work, we demonstrate a new unsupervised deep learning framework using  Convolutional Long-Short Term Memory Network Autoencoders to encode generalized feature representations across both spatial and spectroscopic dimensions spanning $19$ optical emission lines (3800A $< \lambda <$ 8000A) among a sample of $\sim 9000$ galaxies from the MaNGA IFS survey. As a demonstrative exercise, we assess our model on a sample of $290$ Active Galactic Nuclei (AGN) and highlight scientifically interesting characteristics of some highly anomalous AGN.  
\end{abstract}

 % We demonstrate the use of our framework as an anomaly detector by quantifying pixel-wise 2D anomaly score maps as a function of different emission line wavelengths and exploring spatial Regions of Interest (ROIs) within each galaxy that have high anomaly scores.

% For many decades, astronomers have relied on integrated galaxy spectra, especially various emission lines occurring across a wide range of wavelengths, to probe fundamental astrophysics of stellar populations and their evolution \citep{Conroy13,Kewley19}. 

\section{Introduction}
\label{sec:introduction}
Galaxy surveys using Integral Field Spectrographs (IFS) have enabled astronomers to go beyond just using traditional integrated spectra and study spatially resolved spectral properties of galaxies and investigate the complex interplay between important physical processes governing galaxy evolution such as star formation \citep{Sanchez20}, active supermassive blackhole at galactic nucleus \citep[AGN;][]{Wylezalek17}, stellar and gas kinematics \citep[e.g.,][]{Cappellari16}. As such, IFS data offers a unique exploratory pathway to study galaxies. However, the large volume and vast dimensionality nature of the data poses as challenge. Thankfully, deep learning has demonstrated to be an effective strategy to parse and learn relevant information within big data.

% However, a key challenge emerges in terms of doing a data-driven exploration and analysis of vast dimensionalities within large volume of data from IFS-based observations. Thankfully, deep learning has demonstrated to be an effective strategy to parse and learn relevant information within big data.

% Such pathways have been especially transformative as AGN-SF connection is currently one of the highly discussed topics in galaxy evolution science \citep[see][]{Wagner16,Cresci18}.

Extracting compressed and generalized feature representations using unsupervised deep learning approaches such as Convolutional Autoencoders (AEs) and Variational Autoencoders (vAEs) has been a common approach in computer vision problems \citep{Rumelhart86,Masci11,Kingma13}. They have also been widely used in astronomy, notably in studying galaxy morphological demographics in imaging data \citep[e.g.,][]{Nishikawa20,Spindler21,Huertas-Company21,Slijepcevic22} and characterization of 1D galaxy spectra \citep[e.g.,][]{Portillo20,Teimoorinia22}.

Recent application driven advancements has led to the use of Long-Short Term Memory Networks \citep[LSTMs;][]{Hochreiter97} to effectively learning complex relationships withing 1D sequential data \citep[e.g.,][]{Sutskever14} and these later have been adopted to sequential imaging data using 2D Convolutional LSTMs \citep[2DConvLSTMs;][]{Shi15}. Marrying the concepts of 2DConvLSTMs with AEs and vAEs, some recent studies have developed frameworks that learn to encode generalized spatio-sequential representations and used them for identifying abnormal samples within data \citep[i.e., anomaly detection; e.g.,][]{Wang18,Yu20}.

A galaxy's spectrum can be conceptually thought of as a 1D sequence and spatially resolved spectral information as a spatially correlated set of sequences. As such, 2DConvLSTMs are a natural methodological fit to learn spatio-spectral relationships within IFS based data, yet they remain to be explored. Motivated by this, we investigate the use of 2DConvLSTMs on spatially-resolved galaxy spectra in an unsupervised, data driven way. We demonstrate its utility to learn spatio-spectroscopic representations and explore different galaxies hosting potentially unusual spectral signatures that can be of high scientific interest.

\section{Data}
\label{sec:data}
%In this paper, we present a new deep learning based representation learning framework to analyze spatially resolved spectroscopic data of $\sim 9000$ galaxies from the MaNGA IFS-based survey. We demonstrate our model's ability to learn compressed embedding of spatio-spectroscopic information and select galaxies hosting unusual and potentially scientifically interesting spectral properties. As a demonstrative exercise, we showcase our model's application on a scientifically relevant sample of MaNGA AGN galaxies from \citealp{Comerford21} (hereafter C20). 
In this section, we describe our sample selection along with the general details of the overall MaNGA galaxy sample,
%and the C20 AGN sample,  
and their MaNGA spectral data cube descriptions. 

\subsection{MaNGA Target Sample \& IFS Data Cubes}
\label{sec:manga_cubes}
Our deep learning model is trained on IFS cubes from the MaNGA sky survey. 
MaNGA is a survey of $\sim10,000$ galaxies taken from the Sloan Digital Sky Survey (SDSS) and observed using an  IFS unit with hexagonally-assembled, 2-arcsec fiber optic cables \citep{Bundy14}. 
%We use the latest and final MaNGA data release (DR17\footnote{https://www.sdss4.org/dr17/manga/}) data release pipeline (DRP) based summary catalog\footnote{https://www.sdss4.org/dr17/manga/manga-data/data-access/} ({\tt version 3.1.1}), which contains IFS observations of $11,273$ objects, among which $10,010$ observations are of unique galaxies with reliable good quality data.
We start with the summary catalog\footnote{https://www.sdss4.org/dr17/manga/manga-data/data-access/} ({\tt version 3.1.1}) from the data reduction pipeline (DRP) on the latest and final MaNGA data release (DR17\footnote{https://www.sdss4.org/dr17/manga/}). This catalog contains IFS observations of $11,273$ objects, among which $10,010$ observations are of unique galaxies with reliable good quality data. We select a subset of $9043$ galaxies that span $z<0.08$ to ensure a consistent spanning of different optical emission lines to be within the wavelength bounds of the MaNGA IFS data used for our training.
 % (see \S\,\ref{sec:data_preperation})

% Briefly, the catalog provides key information about each galaxy including -- a unique identifier ({\tt PLATEIFU}), {\tt RA} and {\tt DEC} of the IFS observation (centered on the target galaxy), and redshift ({\tt z}). 

% In Figure XX, we show the stellar mass vs redshift of parent sample of MaNGA observed galaxies and highlight those chosen by our selection. 

% \begin{figure}[h!]
%     \centering
%     \includegraphics[width=\columnwidth]{icml2023/Manga_Sample_Description.png}
%     \caption{An overview of the MaNGA DR17 sample in redshift vs. stellar-mass space (black $+$ gray points) shown along with our selection cut (vertical line) and the $406$ MaNGA AGN selected by \citep[][red points]{Comerford21}. }
%     \label{fig:manga_sample}
% \end{figure}

For the $9043$ selected galaxies, we query and download their corresponding DRP reduced IFS spectral data cubes\footnote{see DRP 3D Output Files section in \url{https://www.sdss4.org/dr17/manga/manga-data/data-access/}}. MaNGA observed their target sample of galaxies with different IFS field-of-view layouts ranging from 19 fibers (diameter $12$") to 127 fibers (diameter $32$"). Each raw downloaded spectral cube is a multi-extension FITS file, which contains the flux information (in units of $\mathrm{10^{-17} erg/s/cm{^2}/Angstrom/spaxel}$) for an observed wavelength range $3600\AA\lesssim \lambda \lesssim 10300\AA$ with logarithmic wavelength spacing. 
%Later in our analysis, we further process these raw spectral cubes to be used for our model training.
In \S\,\ref{sec:data_preperation}, we detail our further processing steps for these raw spectral cubes.

\subsection{MaNGA AGN-Hosting Galaxies}
\label{sec:comerford_sample}
To showcase the working of our framework trained on the aforementioned MaNGA sample, we assess a sub-sample of $290$ AGN galaxies within MaNGA selected from an overall sample of $406$ AGN \citep[][hereafter, C20]{Comerford21} after imposing a redshift selection cut $z<0.08$. Briefly, these AGN galaxies have been selected using a wide variety of selection criteria (as indicated with specific flags in the catalog) -- mid-IR WISE colors, hard X-rays (from SWIFT/BAT), NVSS and FIRST radio observations, and presence of broad emission lines in SDSS spectra. These galaxies span a wide range of physical properties including spiral morphology dominant, star-forming main sequence and quiescent elliptical galaxies (see Figure 3 from \citetalias{Comerford21}). 

\begin{figure}[h!]
    \centering
    \includegraphics[width=\columnwidth]{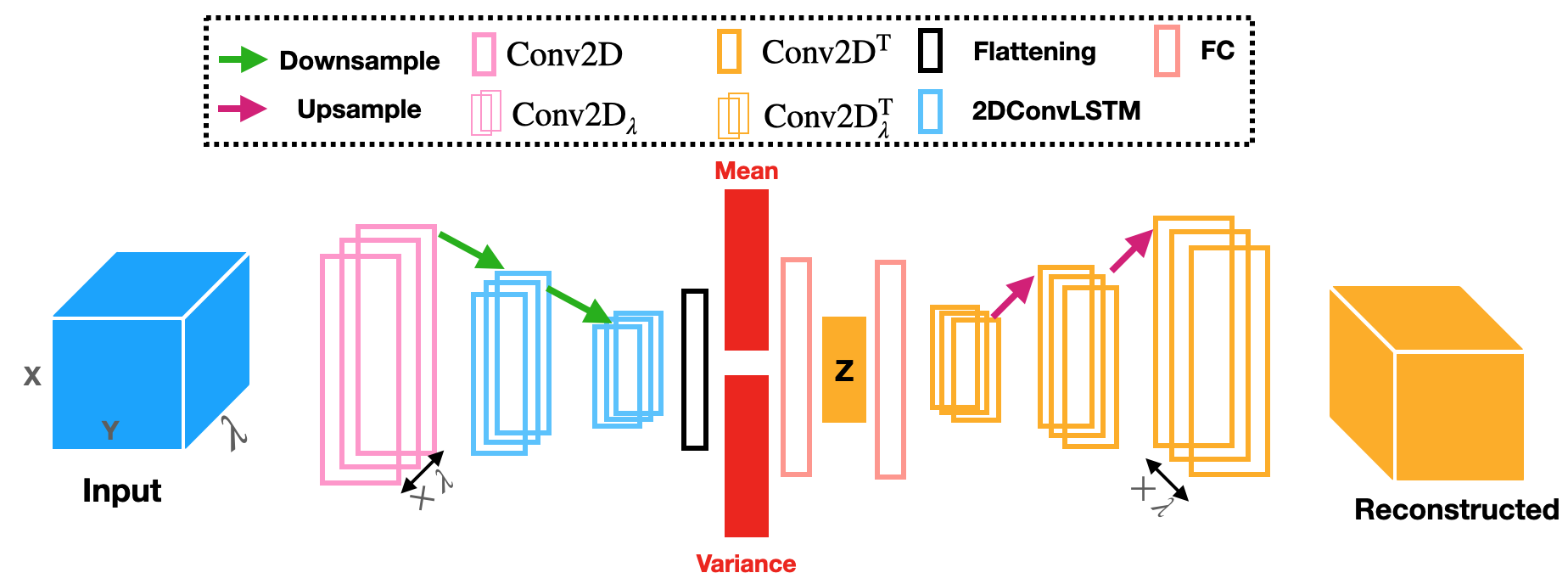}
    \caption{A conceptual overview of our 2DConvLSTM-vAE framework. For the 2DConvLSTM-AE architecture, the mean and variance layers are omitted.}
    \label{fig:architecture}
\end{figure}

 % we develop and apply a new 3D unsupervised deep learning framework, which use the concepts of Long-Short Term Memory Networks (LSTMs) and Autoencoders (AEs) to learn spatial and spectroscopic feature representations from the IFS-based galaxy spectral cubes, and use them to perform anomaly identification.

\begin{figure*}[h!]
    \centering
    \includegraphics[width=1.5\columnwidth]{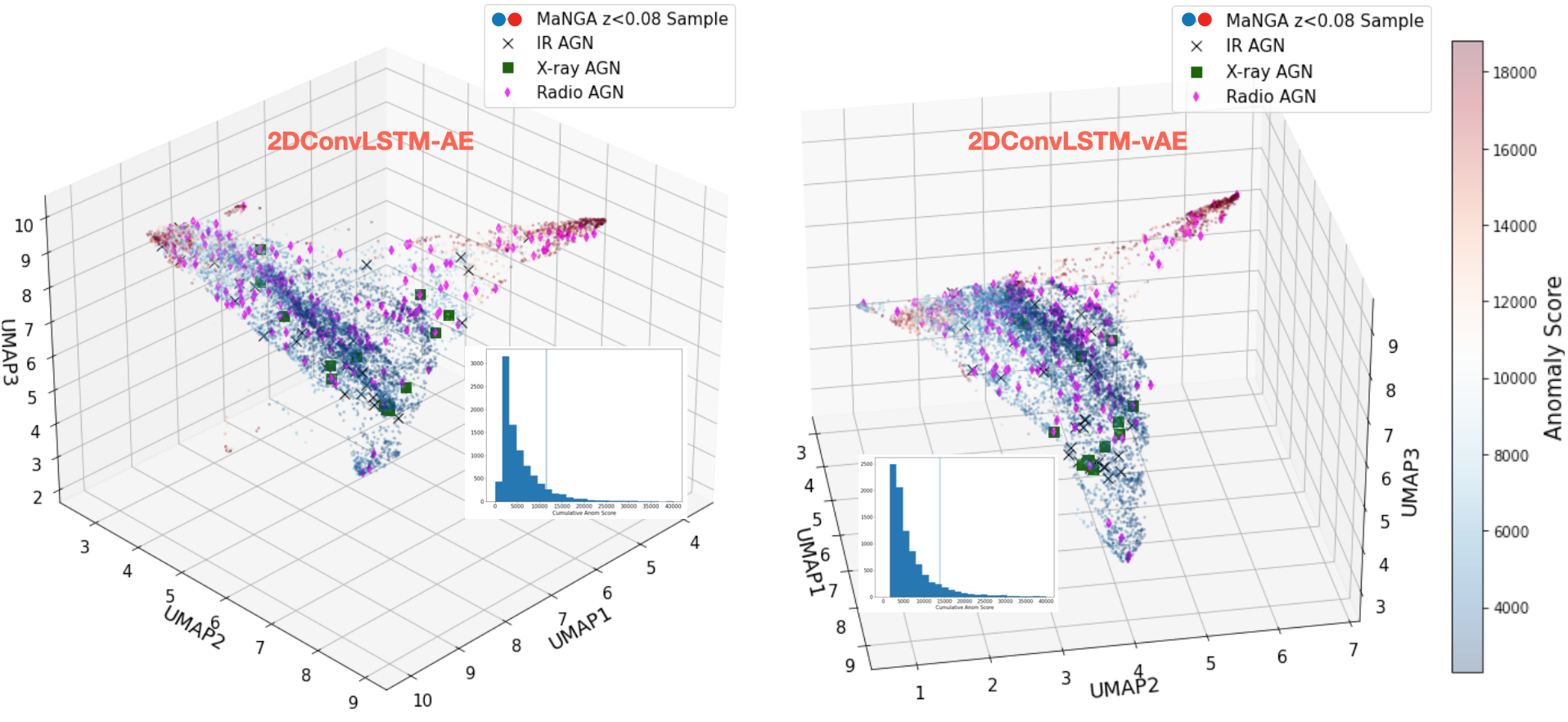}
    \caption{Visualization of latent space embedding for our $\sim 9000$ galaxies extracted from the 2DConvLSTM-AE (left) and 2DConvLSTM-vAE (right) using a UMAP along with the inset plot showing the histogram of anomaly scores.}
    \label{fig:latent_space_anom_score_AE}
\end{figure*}

\section{Deep Learning Model Framework}
\label{sec:framework}
We develop and apply two model architectures (using {\tt Tensorflow}) -- i) a 2D Convolutional LSTM Autoencoder (2DConvLSTM-AE); ii) a Convolutional LSTM variational Autoencoder (2DConvLSTM-vAE), where they follow the standard (Encoder-Bottleneck-Decoder) architecture style of Convolutional AE and vAEs in the literature (see Figure\,\ref{fig:architecture} for a conceptual overview). 

In the encoder stage, both 2DConvLSTM-AE and 2DConvLSTM-vAE frameworks start with an input 3D spectral cube ($X\times Y \times \lambda$) which is then passed to two sets of wavelength-wise 2D Convolutional blocks ({\tt Conv2D$_{\lambda}$)}, where each {\tt Conv2D$_{\lambda}$} block is a series of $\lambda$ {\tt Conv2D} layers acting on 2D $X\times Y$ slices. Following this are three 2D Convolutional LSTM ({\tt 2DConvLSTM}) blocks (with a downsampling factor of 2 in each block). In the case of a 2DConvLSTM-AE, the output of previous block is flattened and then passed to a series of three fully-connected ({\tt FC}) layers, where the central-most {\tt FC} layer is considered to be the ``bottleneck'' ({\tt Z}) with dimensions equaling the latent vector embedding size of choice. In the case of the 2DConvLSTM-vAE model, the flattened set of features are passed to two {\tt FC} layers that describe the mean ($\mu_{z}$) and log-variance ($\log \sigma_{z}^{2})$ of the multi-dimensional latent space distribution. The latent vector ({\tt Z}) is drawn randomly from a Gaussian distribution with that $\mu_{z}$ and $\sigma_{z}$. For both our models, we use $512$ dimensions as the {\tt Z} vector embedding size.

% which act as 2D spatial feature extractors at different wavelengths. These features are then passed as input to a series of three 2D Convolutional LSTM ({\tt 2DConvLSTM}) blocks (with a downsampling factor of 2 in each block), where they further extract spatial features while also simultaneously learning their correlations across the wavelength dimension. In the case of a 2DConvLSTM-AE, the set of extracted spatio-spectroscopic features are first flattened and then passed to a series of three fully-connected ({\tt FC}) layers, where the central-most {\tt FC} layer is considered to be the ``bottleneck'' ({\tt Z}) with dimensions equaling the latent vector embedding size of choice. In the case of the 2DConvLSTM-vAE model, the flattened set of features are passed to two {\tt FC} layers that are described by the mean ($\mu_{z}$) and log-variance ($log \sigma_{z}^{2})$ of the multi-dimensional latent space distribution, and the ({\tt Z}) vector is drawn randomly from a Gaussian distribution with that $\mu_{z}$ and $\sigma_{z}$. For both our models, we use $512$ dimensions as the {\tt Z} vector embedding size.

During the decoder stage, both 2DConvLSTM-AE and 2DConvLSTM-vAE models follow the same structure. First, the latent space vector ({\tt Z}) is repeated $\lambda$ times and is reshaped to ensure consistency to the encoder output dimensionality. This is then passed to a series of three wavelength-distributed 2D Transpose Convolutional blocks ({\tt Conv2D$^{T}_{\lambda}$)}, where each {\tt Conv2D$^{\rm T}_{\lambda}$} consists of  $\lambda$ {\tt Conv2D$^{\rm T}$} layers that progressively upsample the features to the output reconstructed image with the same dimensionality as the input cube. In both the encoder and decoder stages, all but the final convolutional block employ a {\tt Linear} activation and are followed by a {\tt Layer Normalization} layer \citep{ba16}. We use {\tt ELU} (exponential linear unit) as our final layer activation.  

% In Figure\,\ref{fig:architecture}, we show the architecture of both 2DConvLSTM-AE and 2DConvLSTM-vAE models.

% Thus far, we have detailed our selected sample of $\sim 9000$ galaxies and their corresponding spatially-resolved spectral cubes from the MaNGA survey, and detailed the architectures of our deep learning frameworks. {\color{red} As motivated earlier, emission lines can encode a wealth of information about the physics that is driving the energetics of the ISM/IGM.} 

\section{Methodology}
\label{sec:methodology}
%In this work, we train our frameworks to specifically learn the feature encoding based on different emission lines. 
In this section, we describe the preparation steps involved in processing the raw MaNGA spectral cubes to generate their corresponding emission-line spectral cubes along with our employed data augmentation strategy. We also describe our training process and employed hyper-parameters. 

\subsection{Data Preparation}
\label{sec:data_preperation}
We describe the construction of emission-line-only spectral cubes that sample the raw MaNGA cubes described in \S\,\ref{sec:manga_cubes} at $19$ different optical emission lines covered in the SDSS survey\footnote{\url{https://classic.sdss.org/dr6/algorithms/linestable.html}} -- starting from the OII doublet (rest-frame $\lambda = 3272\,\AA$) to the SII doublet (rest-frame $\lambda = 6733\,\AA$). We chose these emission lines to probe key aspects of the physics in galaxies (e.g., star formation, AGN).  For each galaxy in our sample, we start by identifying the observed frame wavelengths corresponding to the $19$ emission lines using its redshift value. Then, by treating the identified observed-frame wavelengths as the centers, for each emission line, we define a wavelength window with bin width of ten, which corresponds to $\Delta\lambda\sim 6\,\AA$. This ensures that our model learns correlations across the emission line profiles rather than a singular central values. The generated emission-line only spectral cubes have $190$ wavelength-wise dimensions.

Owing to the different number of fibers used for observing different galaxies in the MaNGA sample, the resultant spectral cubes span different spatial spaxel dimensions (ranging between $32\times32$ to $127\times 127$). To ensure consistent dimensions for model training, we crop each emission-line cube to  ($32\times32$) with the galaxy at its center (using {\tt RA} and {\tt DEC} information). Each cropped emission-line spectral cube has ($32 \times 32 \times 190$) dimensionality, which are the inputs to our deep learning frameworks. We do not normalize the flux amplitudes of our spectral cube data and train our models to learn in native flux value space. 

% In Figure\,\ref{fig:emline_cube_process}, we visually show the conceptual process of generating the emission-line spectral cube.

Furthermore, we apply a data augmentation strategy to our emission-line cubes for the models to generalize the feature representation better with respect to the different spatial and morphological variances. We augment each cube three times to randomly incorporate a combination of the following transformations -- horizontal flip, rotation by 90 degrees, Gaussian noise, and random translation along the spatial axes by up to 5 pixels. Therefore, we have $\sim 36,000$ augmented emission-line spectral cubes for our model training.

% \begin{figure}[h!]
%     \centering
%     \includegraphics[width=\columnwidth]{icml2023/EM_cube_process.png}
%     \caption{Our data processing steps to generate emission-line only spectral cubes spanning $19$ different optical emission lines between $3272\,\AA \leq \lambda \leq 6733\,\AA$. }
%     \label{fig:emline_cube_process}
% \end{figure}

\subsection{Training Strategy}
\label{sec:training}
We follow standard procedures to train our models. First, we shuffle our augmented spectral cube data set and randomly split the sample into $\sim 30,000$ training and $\sim 6000$ test samples. Next, for our 2DLSTM-AE model, we define our loss function as:
\begin{equation}
    \mathcal{L_{\rm rec}} = \overline{ \left \{ \sum_{x,y} \left [ MAE(I,I^{\prime}) \right ]_{\lambda} + \sum_{\lambda} \left [ MAE(I,I^{\prime}) \right ]_{x,y} \right \}}^{\rm\, batch},
\end{equation}
where $I$ and $I^{\prime}$ are the input and the reconstructed spectral cubes, respectively. In the case of our 2DLSTM-vAE, we minimize a summation of the $\mathcal{L_{\rm rec}}$ and the Kullback-Leibler (KL) Divergence loss between the latent vector (with $\mu_{z}$ and $\sigma_{z}$) and a unit normal distribution. We trained our models with a batch size of 16 cubes and a learning rate of $0.01$ and an Adaptive gradient ({\tt Adagrad}) optimizer to convergence (30 epochs).

% We trained two deep learning frameworks (2DLSTM-AE and 2DLSTM-vAE) on a sample of $9043$ spatial-resolved IFS emission line galaxy spectra from the MaNGA survey.
\section{Results \& Discussion}
\label{sec:discussion}
We analyze the latent space embedding from our 2DLSTM-AE and 2DLSTM-vAE models along with a per-galaxy metric that serves as a proxy for how unusual/anomalous they are. We showcase and discuss the characteristics of highly anomalous galaxies from the general MaNGA sample and also the C20 AGN. We probe the latent space using a nearest neighbour similarity search with the highly anomalous AGN as anchor points and assess the yielded candidates. 

% some AGN recovered within the C20 sample and also some candidates that are not included in C20 that are indicative of hosting an AGN.
% Using our trained 2DLSTM-AE and 2DLSTM-vAE models,

\subsection{Latent Space \& Anomaly Score Distribution}
\label{sec:latent_space_and_anomalies}
For each galaxy, we extract its latent vector from both the models, and compute the mean absolute reconstruction error (MAE) between the input and output reconstructed cubes (hereafter referred to as ``anomaly score''). We then extract the first 50 principle components of our sample $Z$ vectors (corresponding to $\sim 90\%$ explained variance) and further process them to be visualized using Uniform Manifold Approximation and Projection \citep[UMAP;][]{McInnes18}. In Figure\,\ref{fig:latent_space_anom_score_AE}, we show the latent space distribution in 3D UMAP space and color code each data point with its corresponding anomaly score value. The median anomaly scores for our sample are $\sim 3000$ ($\sim 5000$) and $90^{\rm th}$ percentile is $\sim 12000$ ($\sim 20000$) for the 2DConvLSTM-AE (2DConvLSTM-vAE), respectively. We note that bulk of our galaxies with low anomaly scores are distributed across moderate-to-high values across the three UMAP dimensions, whereas the those with high scores span the ``wings'' of the spatial toplogy with low UMAP1 and UMAP2 and preferentially high UMAP3 values. 

In Figure\,\ref{fig:latent_space_anom_score_AE}, we also show the latent space distribution of the $290$ \citetalias{Comerford21} AGNs, split into subgroups that indicate which selection process they were chosen from (Infrared, X-ray, Radio). Generally, we notice that most AGN fall within or close to the loci of low anomaly score samples, although some subset of them have high anomaly scores and fall within the wings of the distribution similarly to other highly anomalous galaxies from our overall sample. While keeping in mind the number dominance of the radio selected galaxies within the \citetalias{Comerford21} sample, we note that most of the anomalous AGN are radio selected, with a handful that are X-ray and IR selected; most X-ray and broad-line selected span different UMAP regions populated predominantly by low to intermediate anomaly score samples. This showcases the identification of anomalous AGN galaxies within IFS data in an unsupervised way.  

\begin{figure}[h!]
    \centering
    \includegraphics[width=\columnwidth]{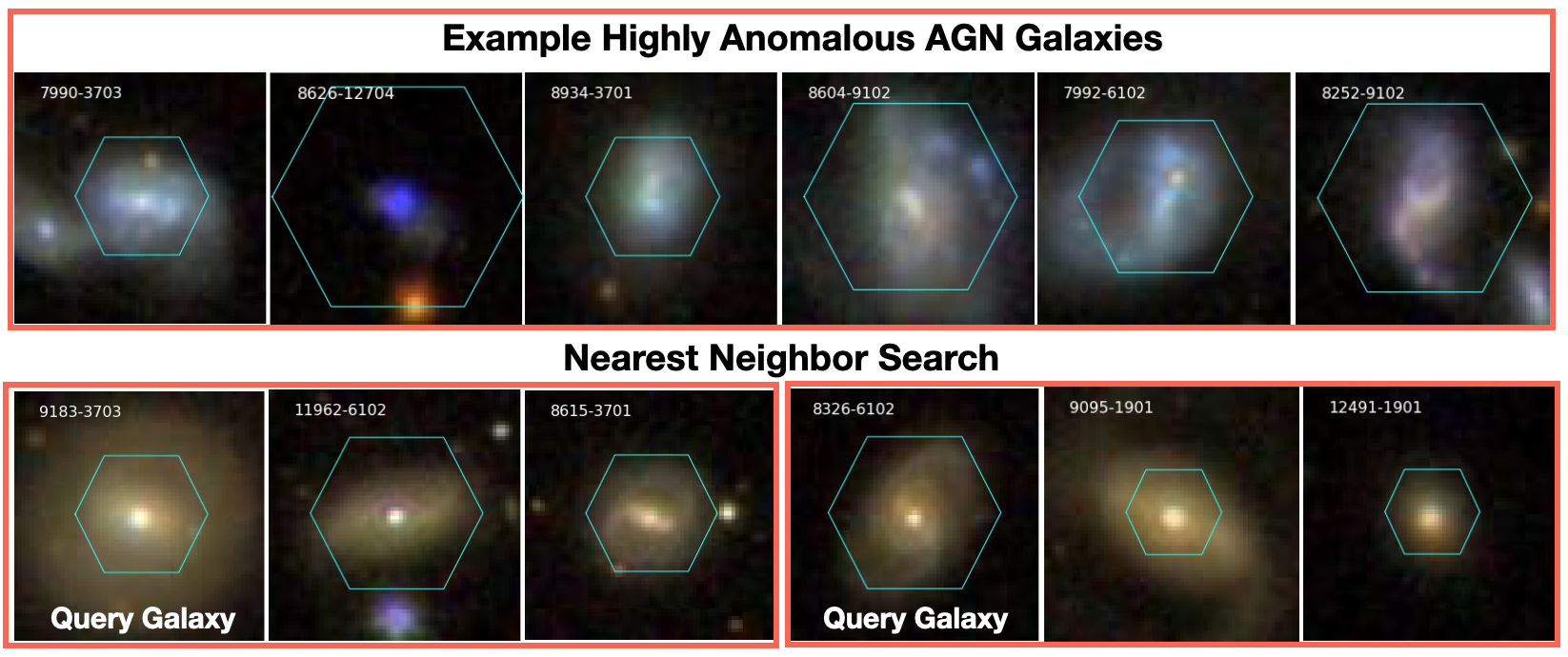}
    \caption{Example highly anomalous AGN galaxies (top panel) and example showcase of latent space nearest neighbour search for two anomalous AGN as query galaxies (bottom panel).}
    \label{fig:nn_search}
\end{figure}

\subsection{Assessment of Highly Anomalous AGN \& Nearest Neighbor Search}
\label{sec:assessment_of_agn}
In Figure\,\ref{fig:nn_search}, we showcase some of the highly anomalous AGN galaxies within our sample. Using the online data exploration and visualization tool {\tt MaNGA Explorer}\footnote{\url{https://manga.voxastro.org/}}, we observe that these highly anomalous AGN spanned a wide range of physical characteristics spanning between having a disturbed morphology, substantial blue/purple emissions as seen in the RGB images indicative of co-existent rapid star formation. Additionally, they have very strong emissions in one ore more emission lines, and contain AGN-like signatures when assessing the provided spaxel-wise BPT diagnostic plots\footnote{([$OIII]/H{\beta}$ vs. $[NII]/H\alpha$ vs. $[SII]/H\alpha$)}. Notably, one of highly anomalous AGN galaxies: {\tt 8626-12704} (second panel in Figure\,\ref{fig:nn_search}), is a ``Blueberry'' galaxy, recently studied as an object of high scientific interest by \cite{Paswan22a,Paswan22b}. 

% \begin{figure}[h!]
%     \centering
%     \includegraphics[width=\columnwidth]{icml2023/Anomalous_AGN.png}
%     \caption{Example highly anomalous AGN galaxies.}
%     \label{fig:anom_agn}
% \end{figure}

To showcase the learned representations and their potential relevance to the physical nature of the galaxies,  we perform a Nearest Neighbour (NN) search around some of the anomalous AGN galaxies. We use the {kd-Tree} method to compute the pair-wise Euclidean distances between different galaxy data points in the UMAP space and find NNs per query galaxy. In Figure\,\ref{fig:nn_search}, we show six example AGN galaxies from C20 with high anomaly scores (top panel) and show NNs for two anomalous AGN (bottom panel). We used {\tt MaNGA Explorer} to investigate the nature of these NNs and find that they indeed show emission line ratios (in the BPT plots) indicative of an AGN.

\section{Conclusions}
\label{conclusions}
In this work, we implemented an unsupervised 2D Convolutional Long-Short Term Memory Autoencoder and Variational Autoencoder to learn feature representations from spatially resolved galaxy spectra information spanning  $19$ optical emission lines (3800A $< \lambda <$ 8000A) among a sample of $9043$ galaxies from the MaNGA survey. We find that our models learn meaningful representations from the data where certain galaxies, including those hosting an AGN are identified to be anomalous and span a wide range of morphological and spectroscopic properties. Using this representation space, we are able to successfully demonstrate our ability to filter and query anomalous galaxies with similar properties.
%Using anomalous AGN as a test case, we demonstrate our models ability to yield back nearest neighbor candidates within the representation space that indeed host AGN signatures. 

% We demonstrate the use of our framework as an anomaly detector by quantifying pixel-wise 2D anomaly score maps as a function of different emission line wavelengths and exploring spatial Regions of Interest (ROIs) within each galaxy that have high anomaly scores. As a demonstrative exercise, we apply our framework on a sample of $406$ Active Galactic Nuclei (AGN) hosting MaNGA galaxies and highlight scientifically interesting characteristics of our model-identified highly anomalous AGN.  

% \section*{Acknowledgements}

% \textbf{Do not} include acknowledgements in the initial version of
% the paper submitted for blind review.

% If a paper is accepted, the final camera-ready version can (and
% probably should) include acknowledgements. In this case, please
% place such acknowledgements in an unnumbered section at the
% end of the paper. Typically, this will include thanks to reviewers
% who gave useful comments, to colleagues who contributed to the ideas,
% and to funding agencies and corporate sponsors that provided financial

% \newpage
\bibliography{example_paper}
\bibliographystyle{icml2023}

%%%%%%%%%%%%%%%%%%%%%%%%%%%%%%%%%%%%%%%%%%%%%%%%%%%%%%%%%%%%%%%%%%%%%%%%%%%%%%%
%%%%%%%%%%%%%%%%%%%%%%%%%%%%%%%%%%%%%%%%%%%%%%%%%%%%%%%%%%%%%%%%%%%%%%%%%%%%%%%
% APPENDIX
%%%%%%%%%%%%%%%%%%%%%%%%%%%%%%%%%%%%%%%%%%%%%%%%%%%%%%%%%%%%%%%%%%%%%%%%%%%%%%%
%%%%%%%%%%%%%%%%%%%%%%%%%%%%%%%%%%%%%%%%%%%%%%%%%%%%%%%%%%%%%%%%%%%%%%%%%%%%%%%
% \newpage
% \appendix
% \onecolumn

% You can have as much text here as you want. The main body must be at most $8$ pages long.
% For the final version, one more page can be added.
% If you want, you can use an appendix like this one, even using the one-column format.
%%%%%%%%%%%%%%%%%%%%%%%%%%%%%%%%%%%%%%%%%%%%%%%%%%%%%%%%%%%%%%%%%%%%%%%%%%%%%%%
%%%%%%%%%%%%%%%%%%%%%%%%%%%%%%%%%%%%%%%%%%%%%%%%%%%%%%%%%%%%%%%%%%%%%%%%%%%%%%%

\end{document}